\documentclass{ws-p9-75x6-50}

\begin{document}

\title{Group Averaging and Refined Algebraic Quantization:  Where are
we now?}

\author{D. Marolf}

\address{Physics Department, Syracuse University, Syracuse, NY 13244 
USA\\E-mail: marolf@phy.syr.edu}


\maketitle

\abstracts{
Refined Algebraic Quantization and Group Averaging are
powerful methods for quantizing constrained systems.  They
give constructive algorithms for generating observables and the physical
inner product.  This work outlines the current status of these ideas
with an eye toward quantum gravity.  The main goal is
provide a description of outstanding problems and possible research topics
in the field.  
}

\section{Introduction}

The well-known
connection between gauge symmetries and first class constraints means
that constrained systems are ubiquitous in modern physics.  Studying
quantization of such systems dates back to Dirac\cite{Dirac}, who
described a general procedure.
However, this `procedure' is far from algorithmic and leaves open
many questions.  While in practice there is little difficulty for
simple systems, issues such as
the construction of both observables and the physical
inner product take on a qualitatively different character for more
complicated systems, such as gravity\cite{KKCI}.  
In the gravitating case, these
issues are associated with the famous `problem of time.'  
Thus, an interest in canonical quantum gravity\cite{KKCI,AAbook} forces
one to reexamine various subtleties. 
We remind the reader that Dirac style methods are common both in
the proposed loop representation\cite{AAbook,CR} for full
quantum gravity and in
quantum cosmology, whether inspired by Einstein-Hilbert 
gravity\cite{ehg,Wald} or string/M-theory\cite{BFM,CU}.  

Refined algebraic quantization\cite{ALMMT,GM1}and related work\cite{AH,KL,QORD}
form an on-going
program to resolve the subtleties of Dirac quantization.  It is closely
associated with a technique known as `group averaging.'  In particular 
it was recently shown\cite{GM2} that, when it is 
well-defined, group averaging gives the unique
implementation of refined
algebraic quantization.  A strength of this approach is that
it gives a constructive method for obtaining both the physical inner
product and interesting observables.  
Other programs
to refine Dirac's scheme include geometric quantization\cite{W,Tu,DE},
BRST\cite{BRST,HT} methods,
Klein-Gordon methods\cite{Wald}, 
coherent state quantization\cite{JK}, C*-algebra methods\cite{ger}, and
algebraic quantization\cite{AAbook}.

The goal of this work is
to outline the current state of refined algebraic quantization
with emphasis on open questions
and possible research projects.  
First, however, we give a brief review of the method.
More complete introductions and reference lists are
contained in other 
works\cite{ALMMT,GM1,E1}.  While we do point out
important technical caveats, we will avoid going into them in detail.  

\section{Review of group averaging and refined algebraic quantization}

As our procedures are based on that of Dirac, we discuss two spaces of
states.  The first is an auxiliary or kinematical space, on which
the constraints are to be formulated as operators.  The goal of the Dirac
scheme is to construct a second space of `physical' states
$|\psi \rangle_{\rm phys}$ which in some sense solve the
constraints.  

The physical states must then be made into a Hilbert 
space. To do so requires an
inner product, which can be particularly
obscure.  Indeed, when the constraints cannot be solved exactly, 
one might wonder how such an inner product could be constructed.  
A natural idea\cite{AAbook} 
is to relate it to properties of the observable algebra.
However, this algebra also typically cannot be found in closed form.

The idea of refined algebraic quantization is to relate construction of
the physical inner product to a much easier problem.  Dirac's space of
auxiliary (or kinematical) states is in some sense
a quantization of a corresponding unconstrained system, as one simply
ignores the constraints at this stage.  This system
can be quantized as usual, without subtleties stemming from constraints.
Thus, if one wished, one make the kinematical states into a Hilbert 
space ${\cal H}_{kin}$ by supplying an   
inner product.
In particular, one may choose the inner product on this
space to map certain real functions on the unconstrained
phase space to Hermitian
operators on ${\cal H}_{kin}$.

Now, given the Hilbert space ${\cal H}_{kin}$, what
group averaging supplies is a constructive method for solving the constraints
and defining a `physical' inner product on the solutions.  This
inner product captures the reality conditions of
observables in the following sense:  Given a Hermitian
operator ${\cal O}$ on ${\cal H}_{kin}$
that commutes with the constraints, it defines a Hermitian operator on the
physical Hilbert space.
Since the classical reality conditions are already
encoded in the inner product on ${\cal H}_{kin}$, this process also
implements them on the physical Hilbert space.
In addition, group averaging gives a method of first constructing 
these observables on ${\cal H}_{kin}$.

Let us turn to the details of these constructions.
It is sufficient to consider the case where the classical constraints
are real.  Furthermore, for the moment we take the
constraints to form the Lie algebra of a unimodular Lie group $G$.      
More general Lie groups will be discussed shortly.  We note however that 
the hypersurface
deformation algebra of canonical gravity
does not form a Lie algebra at all.  Instead of structure constants, 
this algebra contains `structure functions.'  Whether group averaging can
be generalized to accommodate structure functions is an interesting
research question to which considerable discussion will be devoted below.

Since the classical constraints are real, it is natural to implement
them on ${\cal H}_{kin}$ as Hermitian operators\footnote{This is in
general not possible when the constraint algebra contains structure 
functions\cite{Karel}, thus the 
obstacle to treating structure functions with group
averaging.} which then
generate a unitary representation $U$ ($g \mapsto U(g)$) of the 
gauge group $G$.  Dirac would ask that the physical states solve
the constraint in the sense that they be annihilated by the constraints.
The same requirement is given by the statement that
the unitary operators $U(g)$ should act trivially on the physical
states for any $g$ in the gauge group:
\be U(g) |\psi
\rangle_{\rm phys} = |\psi \rangle_{\rm phys}.  
\ee
Now, as 1 need not lie in the discrete spectrum of $U(g)$,  
the Hilbert space ${\cal H}_{\rm kin}$ need not contain any
such solutions.  This is the typical behavior for non-compact groups.
However, we may seek solutions in a suitable space $\Phi^* \supset {\cal H}_{kin}$
of `generalized
states.'  One chooses a subspace $\Phi \subset {\cal
H}_{\rm kin}$ of `test states' and then takes 
$\Phi^* \supset {\cal H}_{kin}$ 
to be the space of all linear functionals on $\Phi$ that
are continuous with respect
to the topology induced from ${\cal H}_{\rm kin}$.  The space
$\Phi$ should be chosen so that the operators $U(g)$ map $\Phi$ into
itself.  In this case, there is a well-defined dual action of $U(g)$
on $f \in \Phi^*$ given by  $[U(g) f](\phi) = f(U(g^{-1})\phi)$ for all
$\phi \in \Phi$.  Solutions of the constraints are then elements
$f\in \Phi^*$ for which $U(g) f = f$ for all $g$.

Consider for the moment the case of compact groups.  Then ${\cal H}_{kin}$
can be decomposed as a direct sum of irreducible representations of
$G$.  In this case,
the integral $\int_G dg U(g)$ gives the operator $P_0$ that
projects onto states in the trivial representation of $G$.
In other words, this operator projects onto just the set of states that
solve the constraint.  Here, $dg$ is the (unique) 
Haar measure on the group.

Let us now consider the integral $\int_G dg U(g)$ for arbitrary
unimodular Lie groups.  In general, Lie groups have two Haar measures: one
invariant under left translations ($g \rightarrow g_0g$)
and one invariant under right translations 
($g \rightarrow gg_0$).   Unimodular groups are those groups
for which these Haar measures coincide.  In this case, the (unique) 
Haar measure is also invariant under the map $g \rightarrow g^{-1}$.
All compact groups are unimodular, as are many non-compact groups.  

When the group is not compact, $\int_G dg U(g)$ will not converge as
an operator on ${\cal H}_{kin}$ and so will not define a projector.
However, physical states need not actually lie in ${\cal H}_{kin}.$
Thus, group averaging might still allow us to `project'
states (at least those lying in $\Phi$)
onto solutions lying in $\Phi^*.$  The map
$\eta: \Phi \rightarrow \Phi^*$, 
$\phi \mapsto \langle \phi | \int_G dg U(g)$
is well-defined when the integral
$\int_G dg \langle \alpha | U(g) | \beta \rangle$ converges for all
$\alpha, \beta$ in $\Phi$.  Let us assume that this is so.  Note
that $\eta$ is anti-linear as is natural for a map from
a space ($\Phi$) to its dual ($\Phi^*$).  Translation
invariance of the Haar measure $dg$ guarantees that $\eta(\phi)$ solves
the constraints in the sense described above.

Group averaging uses the image of $\eta$ and the inner product
$(\eta(\alpha),\eta(\beta))_{phys} =  \eta(\beta)[\alpha]$ on this
space to define the physical Hilbert. 
The reversal of positions of $\alpha, \beta$ on the left and right is
due to the anti-linearity of $\eta$.  That this inner product is
Hermitian follows from the invariance of $dg$ under $g \mapsto
g^{-1}.$  If this is to be a valid physical inner product, it must also
be positive definite.  This is true in every case known to the author, but
has not been established in complete generality.  We will return to
the positivity issue below in our discussion of open questions.

The physical inner product has an important property advertised above:
Hermitian gauge invariant observables on ${\cal H}_{\rm kin}$ will become
Hermitian operators on the physical Hilbert space.  
In refined algebraic quantization, observables are required to
include $\Phi$ in their domain and to map $\Phi$ to itself.  Such
operators act on states $\tilde phi$ in the dual space
$\Phi^*$ through  $({\cal O} \tilde \phi)[\alpha] = \tilde 
\phi( {\cal O}^\dagger \alpha)$. 
`Gauge invariance' of such an operator ${\cal O}$ then means that
${\cal O}$ commutes with the G-action on the domain $\Phi$:
${\cal O} U(g) |\phi \rangle=U(g) {\cal O} |\phi \rangle$ for all
$g\in G$, $\phi\in\Phi$.  Any such observable will commute with the
group averaging map $\eta$, in the sense that ${\cal O} \eta(\phi)
= \eta(\cal O \phi)$.  This in turn means that any relation of the form
$A =  B^\dagger$ between two observables $A$ and $B$ on ${\cal H}_{kin}$
also holds between the corresponding observables on ${\cal H}_{phys}.$
This property is what makes the group averaging inner product 
physically interesting.  

Refined algebraic quantization generalizes
this idea to allow any map $\eta$ (called the `rigging map')
 having the properties deduced for group averaging above.
The choice of this map appears to be
an extra degree of freedom.  
$\Phi$ has been chosen.   One can then show
If the group averaging map converges on $\Phi$ 
and yields a nontrivial result, then\cite{GM2} 
(up to an overall scale) it
is in fact the unique rigging map. 

This completes the construction for the unimodular case, but what of
the non-unimodular case?  Now there are two Haar measures: a left
measure $d_Lg$ invariant under $g \mapsto g_0 g$ and a right
measure $d_Rg$ invariant under $g \mapsto gg_0$.  Neither is
invariant under $g \mapsto g^{-1}.$  Convergence
of group averaging (in both measures) now implies\cite{GM2} 
that {\it no} rigging map exists which
i) is Hermitian, ii) commutes with the observables, and iii) solves the 
constraints in the sense that $\eta(U(g)\phi)= \eta(\phi)$.  

However, this is not the end of the story.  One
can use consistency of group averaging\cite{GM2} or geometric
quantization\cite{DE} to argue that, for non-unimodular groups, 
the physical states should not in fact be annihilated by the 
constraints. 
Instead, one finds that physical states $\psi_{phys}$
should satisfy $U(g) \psi_{phys}
= \Delta(g) \tilde \psi_{phys}$, where $\Delta(g)$ is the so-called `modular
function' on the group, a one-dimensional non-unitary representation
of the group, $\Delta: G \rightarrow {\bf R}^+$ satisfying
$d_Rg = \Delta(g) d_L g.$  In terms of the Hermitian generators $C_i$, 
this means that physical states satisfy
\be
\label{nonUc}
C_i |\psi\rangle_{phys} = \frac{i}{2} tr_{ad}(C_i) |\psi \rangle_{phys},
\ee
instead of being annihilated by the constraint.  Here $tr_{ad}$
denotes the trace in the adjoint representation.  
Since $C_i$ is Hermitian, $i tr_{ad}(C_i)$ cannot lie in its spectrum
unless $tr_{ad}(C_i)$ vanishes.  Nonetheless, generalized states $\psi_{phys}
\in \Phi^*$
satisfying (\ref{nonUc}) may exist.
With the understanding  (\ref{nonUc}), convergence
of group averaging guarantees\cite{GM2} that 
$\phi \mapsto \langle \phi | \int_G dg_0 U(g)$ is the unique rigging map. 
Here, $d_0g = \Delta^{1/2} d_Lg = \Delta^{-1/2} d_Rg$ 
is in fact invariant under $g \mapsto g^{-1}$.

Before moving on to open questions,
we comment briefly on constructing observables.
Just as group averaging constructs 
a physical state $\eta(\phi)$ from a kinematical state $\phi$, so
it also constructs observables (acting
on ${\cal H}_{phys}$) from
operators on ${\cal H}_{kin}$ (`kinematical observables').  
The idea is simple.  Given an operator $O$ on
${\cal H}_{kin}$ and two states $\alpha, \beta \in {\cal H}_{kin}$ we
may compute:
$\int d_L g \langle \alpha | U(g) O U(g^{-1}) |\beta \rangle$.
If this expression converges sufficiently rapidly, it defines an observable.
This observation was used\cite{QORD,BIX} in a minisuperspace context to
construct observables of the `evolving constants of motion' type\cite{CR2,BD}.  There, it
was possible to work with such quantum observables (and even to compute
their matrix elements), without being able to write the corresponding
classical observables as explicit phase space functions.

\section{Where are we now?}

In specialized fields, 
it can be difficult for non-experts
to locate the frontier of research.
Newcomers and beginning graduate students are often unsure just
which questions have been answered and which remain open.  In order to
assist such researchers, the rest of this work
outlines open
questions and possible directions for 
investigation in group averaging and refined algebraic quantization.   
Several of associated research projects are straightforward and some
of them are at a level accessible to beginning graduate students.
It is hoped that this list will encourage a broader range
of physicists and mathematicians to contribute to the field.

\subsection{Comparison with other methods}
\label{other}

Direct application of the Dirac quantization scheme has been a favorite
approach to constrained system quantization within much of
the general relativity
community. However, physicists in other fields are more familiar with
other techniques for quantizing gauge systems.  Primary among these
are the Fadeev-Popov method and its extension through BRST\cite{BRST}
quantization.  A careful comparison of group averaging techniques
with these methods is long overdue.  While at some rough level it is known
that all of these techniques are equivalent to Dirac quantization, 
the application of group averaging represents a refinement of the Dirac idea
and one would like to compare the approaches at this finer level of detail.
For example, recalling that these other schemes are often discussed
within a path integral framework and assuming that there is an equivalence, 
one would like to know just which of choice of measure makes the
procedures agree.

A few results relating group averaging to other approaches are known.   
First consider the usual minisuperspace
setting, with a single constraint and a Hamiltonian that vanishes on the
constraint surface.  Such systems may be explored\cite{PI} by writing
the $U(g)$ that appears in the group averaging expression as a path integral.
In that context, one can show that a path integral of the usual Fadeev-Popov
form computes exactly the matrix elements of the group averaging map $\eta$.
That is to say that, given states $\alpha$ and $\beta$ as the boundary
conditions of the path integral, the path integral computes 
$\eta(\beta)[\alpha]$.
By the usual reasoning, this path integral can be written in
the BRST form as well.  One
obtains a specific measure and range of integration over
the Lagrange multipliers.  Some affects of these
details may be seen through semiclassical approximation. 
In particular\cite{PI}, they 
affect how instantons with negative Euclidean action should contribute.  The
conclusion is that Euclidean 
instantons contribute to matrix elements of the rigging map
as $\exp(-|S_E|)$, with a corresponding result for complex instantons.

Another straightforward case occurs when the gauge group
is unimodular and the constraints are linear in momenta.  Note that the term
unimodular necessarily implies that we consider a finite dimensional
Lie group, and in particular that we are in a quantum mechanical setting
as opposed to that of a local field theory.  In this case one
can again introduce a path integral expression for the rigging map and
show that matrix elements of the form $( e^{iHt} \eta (\beta))[\alpha]$
take the form of a standard Fadeev-Popov path integral.  One can also
show directly using operator methods that group averaging, meaning
a scheme using both the group averaging inner product and a construction
of observables through group averaging, is equivalent to a scheme in which
one first fixes a (coordinate) gauge and then quantizes the system.
This is closely related to the observations of Woodard\cite{RW}.
The particularly simple case of the parametrized particle is treated in
the literature,\cite{QORD} 
but the general case of this sort is quite similar.

We see, however, that there are ways in which additional subtleties
can enter the story.  The first is the case of non-unimodular groups.
Due to the usual emphasis on compact (local) gauge groups, these are not typically
considered in the framework of Fadeev-Popov and BRST.  Based on the group
averaging\cite{GM2} and geometric quantization results\cite{DE}
described above, one suspects that Fadeev-Popov and BRST techniques require
a small adjustment in non-unimodular case. After this
is done, one expects to again
find agreement with
group averaging for constraints linear in momenta though this has
not yet been shown.

The second place where subtleties may arise is for
constraints not linear in momenta.  Let us recall first
how this complicates the Fadeev-Popov framework itself.
Suppose that one works in the canonical setting.  Because the constraints
are non-linear in momenta, the Fadeev-Popov determinant will typically
involve momenta as well.  Thus the fully gauge-fixed action
is no longer precisely quadratic in momenta and the translation back and
forth between (gauge-fixed) Lagrangian and Hamiltonian settings is
complicated.  Naive application of Fadeev-Popov methods
in the Lagrangian and Hamiltonian frameworks then leads to inequivalent results.
More precisely, choosing a simple measure for the path integral on one
side may correspond to choosing a complicated measure on the other side.
It is thus of interest to find out what prescription if any corresponds
to group averaging.  Because group averaging is essentially a canonical
scheme, one would expect the measure to appear simple in terms of a canonical
path integral.  This is just what was found in\cite{PI} for the case
of a single constraint, but the general case remains to be done.  
It would also be interesting to revisit Woodard's arguments\cite{RW} with
this in mind.

A field theory setting might provide a third source of subtleties.
Rigorous results for this case may be difficult, but even a heuristic treatment
may prove useful. 
Finally, one would like to investigate further the relation
between group averaging and geometric quantization\cite{W,Tu,DE},
coherent state quantization\cite{JK}, C*-algebra methods\cite{ger}, and
Klein-Gordon methods\cite{Wald} (in the context of constraints quadratic
in momenta).  Some remarks on the relation to
Klein-Gordon methods have appeared\cite{emb,deco,QORD}.  These point out
that one may use the relation between group averaging and Klein-Gordon methods
to define a vacuum state on curved spacetimes, though it is not clear
what physical status this vacuum might have.

\subsection{Structure Functions}

Perhaps the most interesting open question is whether group averaging can
be generalized to systems with structure functions.  Since
structure functions appear in 
the hypersurface deformation algebra
of gravitating systems\footnote{Though they are absent in
minisuperspace models and can be removed in the 1+1 case.},  
this issue must be resolved before one may use group averaging
for full quantum gravity. 

While the leap to structure functions may seem like a large one, 
reasons for optimism exist.  For example, recall\cite{Karel} that 
the difficulty is tied to to the current 
insistence on using Hermitian operators for the constraints.
But perhaps the constraints could have 
an anti-Hermitian part?  Returning to
the case of non-unimodular Lie groups, we see that the answer is
affirmative.  Recall\cite{DE,GM2} that quantization of
non-unimodular Lie groups can proceed with Hermitian constraints $C_i$, but
that physical states satisfy (\ref{nonUc}) instead of the more
familiar condition that they are annihilated by the constraints.
However, (\ref{nonUc}) is equivalent to considering constraints
$\tilde C_i = C_i - i tr_{ad}(C_i)$ which annihilate the constraints.
One may check that the $\tilde C_i$ generate a non-unitary representation
of the same group as the constraints $C_i$.  In fact, if $C_i$ generate
the representation $U(g)$, then $\tilde C_i$ generate $\Delta^{-1}(g)U(g).$
Something similar should arise for structure functions.

Another useful observation is that one may
convert an algebra with structure constants into an algebra with
structure functions by merely multiplying each classical constraint 
by a function on phase space.  A similar statement is true in the quantum
setting if one multiplies the constraints on the left by
operators.  If the original constraints of the Lie algebra
were Hermitian, the new constraints will in general not be so. 
One may hope to gain control over the structure function case by
examining in detail such cases of `artificial' structure functions where
an honest Lie group lurks in the background.
One would hope in this way to build on lessons from
the non-unimodular case and come to terms with structure functions.

\subsection{When Group Averaging Fails}

We have said that group averaging and refined algebraic quantization can
to a large extent be identified.  This follows from a 
theorem\cite{GM2} which states that, when group averaging converges, it
gives the unique implementation of refined algebraic quantization.
However, what happens when group averaging fails to converge is an
interesting and very open area of investigation.

Only a few examples have been studied: the
diffeomorphism constraints of gravity in the loop  
representation\cite{ALMMT}, cases involving 
a single constraint\cite{Ban}, and the action of
SO(n,1) on functions on Minkowski space\cite{GoMa}.
In all cases some `renormalization' of the group averaging map was used.
In particular, a subspace $\Phi$ of the kinematical 
Hilbert space split into different `sectors' in which group averaging
required different amounts of renormalization.   These sectors were
in fact superselected, in the sense that any operator which commutes
with the constraints and preserves $\Phi$ has vanishing matrix elements
between states in different sectors.  This allows a freedom to rescale
the physical inner product separately on each superselected 
sector.  Refined algebraic quantization is no longer unique, though 
this non-uniqueness does not affect any physics.  It is appealing to
think that this is the general
picture, but the evidence is 
too preliminary to draw firm conclusions.  Further investigation
of examples is needed, which one hopes will lead to general
theorems.
    
\subsection{Study of Examples}

Much of the work to date in group averaging and refined algebraic
quantization has centered on deriving general results and developing the
overall structure of the approach.  While this has resulted in powerful
theorems, the methods are sufficiently abstract that it would be useful
to have more concrete examples worked out in detail.  A few examples
were studied in early works\cite{QORD,BIX},  but even there
the detailed answers to physical questions were not computed.  The
focus was on showing that both the physical inner product and operators of the
evolving constant of motion type were well defined.  Missing
were computations of expectation values or matrix elements of such operators in particular
physical states which might shed light on the
physical implications of quantum gravity, and more detailed 
studies of the time evolution inherent in quantum mechanical
evolving constants of motion.  A semi-classical treatment\cite{BDT} is useful
to illustrate the sort of effects that may arise.  

A particularly interesting case is the Bianchi
IX model.  This model has been a favorite of minisuperspace workers
and is now thought\cite{sing} to be connected to the generic behavior of general
relativity near a singularity.  
It has been argued using a WKB-like self-consistent approximation
scheme that refined algebraic quantization can be applied to this model\cite{BIX}
though again without calculations of detailed properties of observables.
This complicated model may be
an appropriate place to introduce numerical techniques, and
the development of numerical techniques for group averaging is an
interesting direction of research in itself.  

In addition, it is by no means clear that the 
constructions
of evolving constants of motion via group averaging
used in the early works\cite{QORD,BIX} are the best.
The goal there was to prove rigorously that a complete
set of such observables exists.  To do so, 
certain ``regulators'' were introduced.  While these regulators have
no effect in the classical limit, they look rather ad hoc from the quantum
mechanical perspective.  One may expect that such regulators are
in fact not necessary, though it will take more detailed calculations 
to show this.  It is exactly this sort of issue
that might be best probed by starting with particular models and working
out the details.

Another feature deserving of more insight is the
$itr_{ad}(C_i)$ term required for non-unimodular groups.
Equation (\ref{nonUc}) implies that the
Hermitian constraints $C_i$ do not annihilate
physical states.  As a result, one might think of
physical states 
as having a certain finite `width' around the constraint surface.  
Since, however, one may also use the 
non-Hermitian
constraints $\tilde C_i$ that do annihilate physical states, it is not clear
whether this is a useful way in which to think about the effect.
Perhaps one can gain more insight into by studying particular systems in detail.

\subsection{Semiclassical Techniques}

Another direction to explore is the use of semiclassical
techniques.   Some results 
about the group averaging map are known\cite{PI}
to leading semi-classical order, $e^{iS/\hbar}$.
However, an extension of this study to higher orders would be worthwhile.
In particular, it is at the next order where one would
expect to see contributions
from the particular measure in a path integral associated with group
averaging.  

A traditional application of semiclassical techniques in quantum gravity
has been to `the problem of time' and further studies are warranted
in the group averaging context.  While much
of the general theory for the minisuperspace case has been developed\cite{BDT},
one would like to see detailed investigations in 
particular models as well as a generalization to
cases with multiple constraints. 

\subsection{Remaining Issues}

A few issues remain that, while they have a rather technical flavor, 
may be of great importance.
The first has to do with an ambiguity
in the approach:  When group averaging converges on a given space $\Phi$, 
one finds\cite{GM2} that this process gives the unique rigging map 
$\eta$.  However, this says nothing about how $\Phi$ is to be chosen
in the first place and to what extent it is unique.   An initial
exploration\cite{Ban} indicated
that $\Phi$ will not be determined by mathematical consistency alone.
Instead, it was conjectured that the choice of $\Phi$ contains 
physical input related to the classical choice of differential
structure on the unconstrained phase space.  In particular, it would
therefore encode the differential structure of the constraint surface.
However, the details of this idea remain to be fleshed out in full.

Another important issue concerns the positivity of the group
averaging inner product.  Suppose that group averaging converges
nontrivially.
Then what is known is that any implementation
of refined algebraic quantization is proportional to the group
averaging result.  Thus, positivity of group averaging is closely
linked to the success of refined algebraic quantization in
constructing a positive definite physical Hilbert space.
No cases are presently known in which group averaging fails to
be positive definite, but this issue clearly deserves further study.

A final issue involves a connection with
representation theory.  Recall our discussion of group averaging for 
compact groups.  There, it yielded
a projector onto a subspace associated with the trivial
representation in a decomposition of ${\cal H}_{kin}$
into irreducibles.  However, this is not the general 
case\cite{GM2} for group averaging.  
Indeed, 
for type II and type III groups\cite{Dixmier} 
(also known as `wild' groups\cite{Ki}), 
a unique decomposition into irreducibles may not exist.  
Now, the usual decomposition into
irreducibles is associated with the mathematical notion of `weak containment'
of representations\cite{Dixmier,Ki}.  One suggestion\cite{GM2} is
that refined algebraic
quantization may instead relate to a new `ultraweak' notion of
containment.  This should be
explored both for the sake of mathematics and with an eye toward understanding
cases where group averaging fails.
As in other settings, a general understanding of the abstract structure
of a procedure may allow one to find and control features which are
difficult to grasp through detailed calculations.
\smallskip

The above `laundry list' of issues shows both the variety of
results obtained to date and the amenability of the field to further
study.  As in many cases, there is a need for both detailed
investigation of particular models and abstract work on
general principles.  Such a field is open 
to a variety of researchers and it is
difficult to predict from which corner the next important
insights will emerge.  

\section*{Acknowledgements}

\smallskip

This work was supported in part by
NSF grant PHY97-22362,
the Alfred P. Sloan foundation, and funds from Syracuse
University.  The author wishes to thank A. Ashtekar, C. Fewster, 
A. Gomberoff, 
A. Higuchi, K. Kuchar, J. Lewandowski,
J. Mour\~ao, C. Rovelli, T. Thiemann, and of course D.
Giulini for many useful
conversations on refined algebraic quantization and group averaging.

\end{document}